\documentclass[aps,prl,superscriptaddress,twocolumn,showpacs]{revtex4}

\renewcommand*{\[}{\begin{equation}}

\renewcommand*{\]}{\end{equation}}

\def\PRA{{Phys.~Rev.~A} }

\def\JPB{{J.~Phys.~B} }
\def\PRL{{Phys.~Rev.~Lett.} }

\newcommand{\myscaleboxb}[1]{\scalebox{0.44}[0.44]{#1}}
\newcommand{\myscaleboxc}[1]{\scalebox{0.58}[0.58]{#1}}
\newcommand{\myscaleboxa}[1]{\scalebox{0.33}[0.33]{#1}}

\usepackage{epsfig}

\usepackage{hyperref}

\begin{document}

\title{A robust all-non-optical method for the characterization of single-shot few-cycle laser pulses}

\author{Zhangjin Chen}

\affiliation{J. R. Macdonald Laboratory, Physics Department, Kansas
State University, Manhattan, Kansas 66506-2604, USA}

\author{T. Wittmann}

\affiliation{Max-Planck-Institue f\"{u}r Quantenoptik,
Hans-Kopfermann-Str. 1, 85748 Garching, Germany}

\author{C. D. Lin }

\affiliation{J. R. Macdonald Laboratory, Physics Department, Kansas
State University, Manhattan, Kansas 66506-2604, USA}

\date{\today}

\begin{abstract}

An all non-optical method for accurately determining the pulse
parameters of individual few-cycle laser shots is presented. By
analyzing the ``left" and ``right" asymmetry of high-energy
photoelectrons along the polarization axis using the recently
developed quantitative rescattering theory, we show that the
carrier-envelope phase, the pulse duration and the peak intensity of
each single-shot pulse can be readily retrieved. Unlike optical
measurements, this method provides the laser intensity in the
interaction region directly.
\end{abstract}


\maketitle

Observation of the interaction of laser pulses with matter requires
the knowledge of the waveform of the laser field. The waveform is
characterized by its pulse envelope, including peak intensity and
pulse duration, as well as the carrier-envelope phase (CEP) which
measures the offset between the peak of the electric field and the
peak of the envelope. Specifically, such a CEP-fixed waveform can be
written as $E(t)=E_0(t)\cos(\omega t+\phi)$, where $\omega$ is the
frequency of the carrier wave, $\phi$ is the CEP.

The stabilization of the CEP is of great importance, for example, in
the generation of isolated attosecond XUV pulses~\cite{kien04}, and
in controlling the dissociation dynamics of
molecules~\cite{kling06}. For atomic targets, the CEP affects the
relative yield of high-energy above-threshold-ionization (HATI)
electrons along the left vs the right directions of the polarization
axis, and the ratio of the left/right photoelectron yield has been
used for the determination of the CEP of few-cycle
pulses~\cite{paulus03}. In such measurements the CEP of a laser beam
must be stabilized. However, the stabilization of the CEP is rather
complex. It is often determined by averaging over the jitters from
many laser shots, thus limiting the temporal resolution of
``CEP-stabilized" pulses. Recently, Wittmann \emph{et
al}.~\cite{Wittmann} reported that the left/right photoelectrons
from individual shots can be measured. In this Letter, we show that
based on the recently developed quantitative rescattering (QRS)
theory~\cite{Chen09}, the CEP, the pulse duration, together with the
peak intensity at the laser focus, can all be accurately determined
simultaneously from the measured electron momentum spectra, thus
providing a complete non-optical method for the full
characterization of each few-cycle laser shot. The method is robust
and the laser pulses can be characterized within a few hours using
typical personal computers.

From the QRS~\cite{Chen09}, the momentum distributions of HATI
electrons can be expressed as
\begin{eqnarray}
\label{QRS}D(p,\theta)= W(p_r)\sigma(p_r, \theta_r)
\end{eqnarray}
where $W(p_r)$ is the returning electron wave packet and
$\sigma(p_r, \theta_r)$ is the elastic differential cross section
(DCS) between the field-free electrons and the target ion. The QRS
is a quantitative version of the well-known rescattering
model~\cite{Krause,Corkum}. In this model, electrons which are
released earlier in the laser field may be driven back to recollide
with the target ion. In Eq.~(\ref{QRS}), the momentum $p_r$ of the
returning wave packet (not directly measured) is related to the
measured photoelectron momentum $p$ via
\begin{eqnarray}
\label{PrAr}p\cos\theta &=& -A_r\mp p_r \cos \theta_r, \\
p\sin\theta &=& p_r\sin \theta_r
\end{eqnarray}
where $A_r$ is the vector potential of the laser field at the time
of recollision. In Eq. (\ref{PrAr}), the upper sign refers to the
right side while the lower sign refers to the left side. (All the
equations are written in atomic units unless otherwise noted.)
Following the QRS~\cite{Chen09, Chenjpb}, the magnitude of
$A_r$=$p_r/1.26$. The angles $\theta$ and $\theta_r$ are defined
with respect to the laser polarization axis, and $\theta_r$ is
measured from the direction of the ``incident beam", i.e., of the
returning electrons. For HATI electrons along the polarization axis,
$\theta$=0 (right) or $\pi$ (left), $\theta_r$=$\pi$, and
$p$=$2.26p_r/1.26$=$1.79 p_r$.

According to the QRS~\cite{Chen09,toru08}, the wave packet $W(p_r)$
depends mostly on the lasers only, while the DCS is solely the
property of the target. Due to this separability the experimental
HATI spectra can also be written as the product of a
volume-integrated ``wave packet" $\overline{W}(p_r)$ with the
DCS~\cite{Chen09}. For each pulse, there are two returning wave
packets, one from the left, the other from the right, along the
polarization axis. Fig.~1(a) shows the ``left" and ``right" electron
spectra from a typical single-shot measurement. Using
$\overline{W}_R(p_r)= D(p,\theta=0)/ \sigma(p_r,\theta_r =\pi)$,
where $\sigma(p_r,\theta_r =\pi)$ for atomic targets are easily
calculated,  for example, the right wave packet can be obtained.
Similarly a left wave packet $\overline{W}_L(p_r)$ can also be
obtained.  By comparing these ``experimental" wave packets with
those $\overline{W}(p_r)$ obtained theoretically, the laser
parameters used in the experiment are retrieved. Since the wave
packet in the QRS is obtained from strong field approximation, the
calculation is a few thousands times faster than from solving the
time-dependent Schr\"{o}dinger equation. This speed-up makes it
possible to carry out near real-time retrieval of laser parameters
from experimental data. To perform volume integration, we assume
that the spatial distribution of each laser shot is Gaussian and all
the electrons from the interaction volume are collected. Thus only
the peak intensity at the laser focus and its pulse length (defined
as the full-width at half-maximum of the intensity), in addition to
the CEP, will be retrieved.  We also assume that the peak intensity
and pulse duration of each individual shot do not change, i.e., only
the CEP is changed randomly. The wavelength is assumed to be known.

In Ref.~\cite{Wittmann}, 4500 single-shot data were collected. We
first determine the peak intensity and pulse duration used in the
experiment. For this purpose, we define a single quantity called
energy moment $M$ for each shot,
\begin{eqnarray}
\label{moment}M=\frac{\int_{p_{r1}}^{p_{r3}}(p_r^2/2)\overline{W}(p_r)dp_r}{\int_{p_{r1}}^{p_{r3}}
\overline{W}(p_r)dp_r}
\end{eqnarray}
where the momentum $p_r$ of the returning electron is related to the
momentum $p$ along the polarization axis (and energy  $E_i=p_i^2/2$)
by $p$=1.79$p_r$. For example, from the experimental electron
spectra like Fig.~1(a) (or better the spectra summed over all the
shots, on the left or on the right) we estimate $E_1$ and $E_3$,
where $E_1$ is close to about 5$U_p$ and $E_3$ is about 10$U_p$,
where  $U_p$ is the ponderomotive energy. These selections are made
since the QRS is valid only for HATI electrons. The precise values
of $E_1$ and $E_3$  are not important. Using the left wave packet
for all the 4500 shots, 4500 values of $M$'s are calculated from
Eq.~(\ref{moment}), using fixed $E_1$ and $E_3$. These calculated
values are displayed in Fig.~1(b) where the horizontal axis is
divided into 90 sections. The moments $M$ calculated from the first
50 shots are placed in the first bin, at the vertical positions
corresponding to the values of $M$. The $M$'s from the next 50 shots
are placed in the second bin. The process continues till the energy
moments from all the shots are registered.  Note that the $M$'s are
distributed nearly uniformly within a band. The average value of
$M$, or $\overline{M}$, was calculated to be 16.46 eV and was found
to be independent of the pulse duration. Using the QRS, the
$\overline{M}$ values  (averaged over the whole 2$\pi$ range of the
CEP) were found  to be 15.64, 16.05 and 16.46 eV, respectively, for
peak intensities of 1.2, 1.3 and $1.4\times10^{14}$ W/cm$^2$. By
choosing peak intensity at $1.4\times10^{14}$ W/cm$^2$, and for
pulse durations of 4.5, 4.7 and 5.0 fs, respectively, we found that
the best fit to the (vertical) band width is for pulse duration of
about 4.6~fs, see Fig.~1(c). Thus the energy moments calculated over
the single shots allow the determination of the peak intensity and
pulse duration easily.

In the above analysis, only the energy moments from the left wave
packets were considered. If the right and left detectors are exactly
identical, then the same peak intensity and pulse duration should be
obtained from the right wave packets. From Figs.~1(c,d), it is clear
that the two detectors are not exactly the same. If we were to use
the data from Fig.~1(d), we would obtain a peak intensity of
$1.33\times10^{14}$ W/cm$^2$ and pulse duration of 4.8~fs. We
checked that these conclusions are not changed much when the values
of $E_1$ and $E_3$ are varied.

\begin{figure}
\mbox{\rotatebox{270}{\myscaleboxb{
\includegraphics{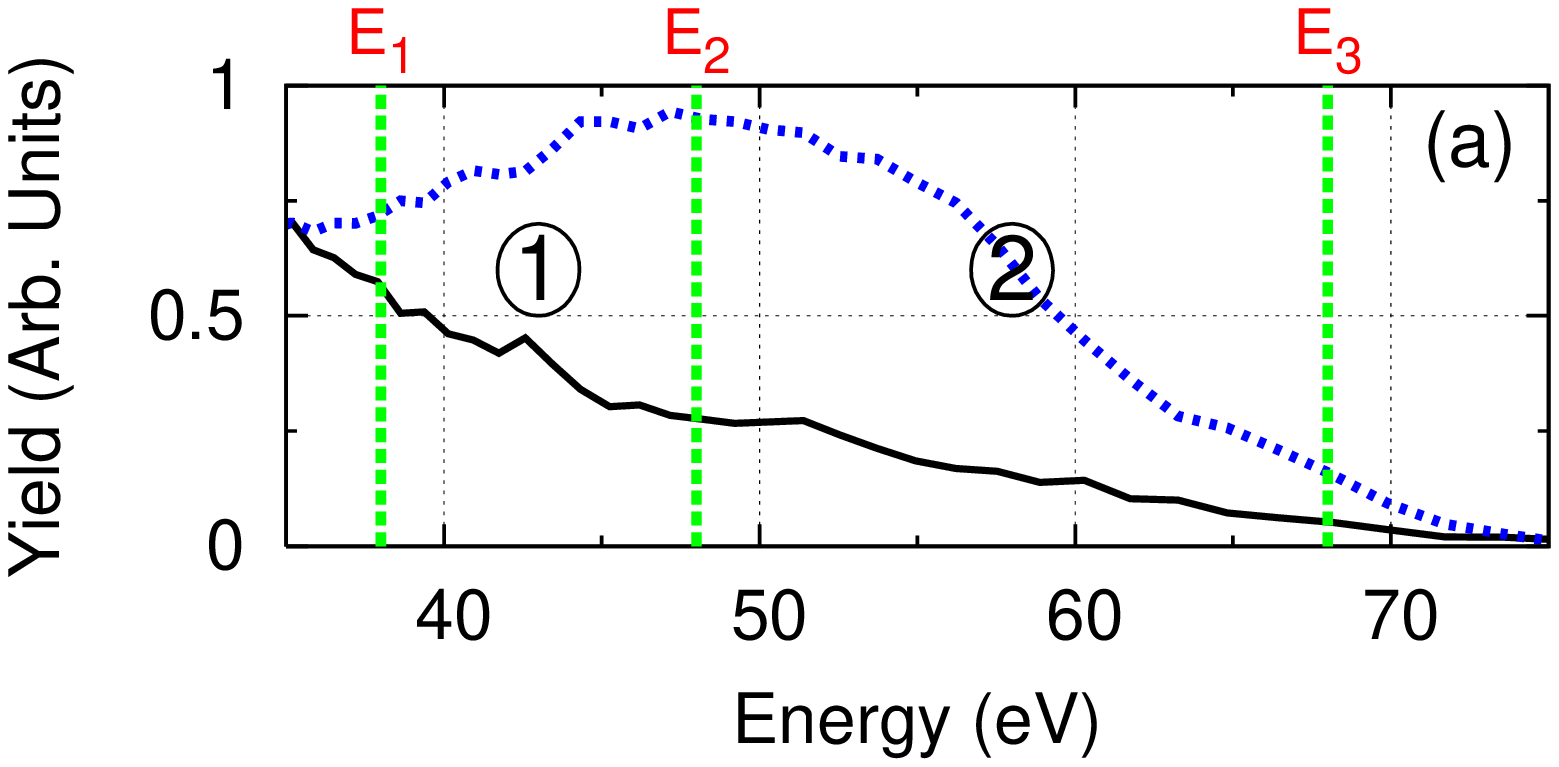}}}}
\mbox{\rotatebox{270}{\myscaleboxb{
\includegraphics{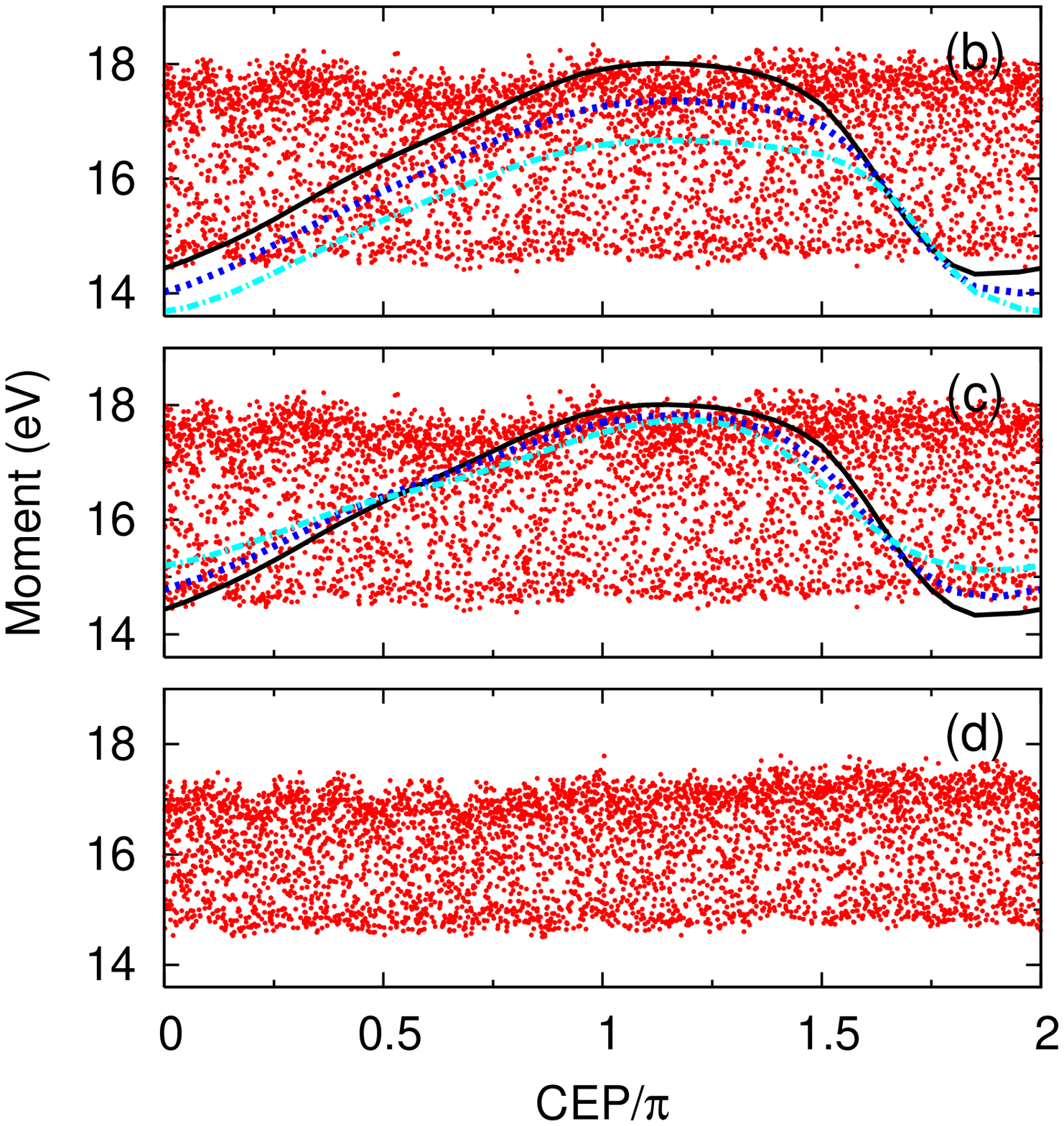}}}}
\caption{(Color online) (a) Typical single-shot left- (solid line)
and right-side (dashed line) electron energy spectra along the
polarization axis. (b) Energy moment of left-side   spectra from
experimental measurements (dots) compared with theoretical
calculations at peak intensities of 1.2 (dash-dot line), 1.3 (broken
line), and $1.4\times10^{14}$ W/cm$^2$ (solid line) with pulse
duration of 4.5~fs; (c) Same as (b) but for theoretical calculations
at peak intensity of $1.4\times10^{14}$ W/cm$^2$, and pulse
durations of 5.0 (dash-dot line), 4.7 (broken line) and 4.5~fs
(solid line), respectively; (d) Energy moment of right-side spectra
from experimental measurements. The experimental data are from
Wittmann \emph{et al}.~\cite{Wittmann} measured for single
ionization of Xe in laser pulses with the wavelength of 760~nm. In
(b-d), experimental data from all 4500 shots are shown, see text.}
\end{figure}

Once the peak intensity and pulse duration are known, we retrieve
the CEP for each shot following the procedure of Wittmann \emph{et
al}.~\cite{Wittmann}. In their method, between $E_1$ and $E_3$,
another intermediate energy $E_2$ was chosen (see Fig.~1(a)). The
total electron yield $Y_L$ between $E_1$ and $E_2$ from the left
detector is evaluated, and a $Y_R$  from the right detector in the
same energy range is calculated. Define the asymmetry
$A_1=(Y_L-Y_R)/(Y_L + Y_R)$. A similar asymmetry parameter $A_2$ is
defined for the electron yields between $E_2$ and $E_3$. Using
($E_1, E_2, E_3$)=(37.9, 57.5, 64.8) eV as in Wittmann \emph{et
al}.~\cite{Wittmann}, the ($A_1, A_2$) for each laser shot is
plotted as a point in 2D, and the results for all the shots are
shown in Fig.~2(a). On top of the plot, three theoretical curves are
shown, for peak intensity of $1.4\times10^{14}$ W/cm$^2$ and pulse
durations of 4.5, 4.7 and 5.0~fs, respectively.  From the three
curves, a duration of 4.7~fs gives the best overall fit to the
experimental data. This number happens to be the average of 4.6 fs
and 4.8~fs derived from Fig.~1. Note that the theory curve is simply
a Lissajous parameter plot of $A_1$ and $A_2$, versus the implicit
variable, the CEP. The theory expects a perfect ellipse. Due to the
intrinsic errors in experimental electron spectra, the experimental
``ellipse" acquires a width, and the ellipse is distorted due to the
difference in the left and right detectors.

\begin{figure}
\mbox{\rotatebox{270}{\myscaleboxc{
\includegraphics{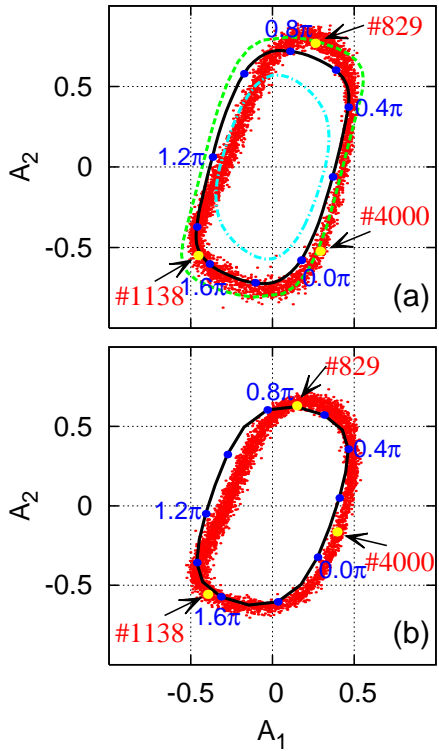}}}}
\caption{(Color online) Comparison of the asymmetry ellipse from
experiment with theory. The experimental   points are calculated
from the data of Wittmann \emph{et al}.~\cite{Wittmann}. The peak
intensity used in the theoretical simulations is $1.4\times 10^{14}$
W/cm$^2$. (a) The energy range used is ($E_1,E_2,E_3$)=(37.9, 57.5,
64.8) eV, and pulse durations   used in theoretical simulations are
4.5 (broken line), 4.7 (solid line) and 5.0~fs (dash-dot line),
respectively; (b) The energy range is ($E_1,E_2,E_3$)=(38.6, 48.7,
82.0) eV and the pulse duration for theory is 4.7~fs (solid line).}
\end{figure}

The shape of the ellipse depends on the choice of ($E_1, E_2, E_3$)
used in the calculation of $A_1$ and $A_2$. In Fig.~2(b), we show
another choice of these parameters. The size and the orientation of
the ellipse are changed. However, the actual retrieved CEP's  are
insensitive to such choices. Take laser shots, \#829, \#1138 and
\#4000 as examples. We retrieve the CEP by drawing a straightline
from ($A_1,A_2$)=($0,0$) to the experimental point (marked by large
yellow dots). From the intercept of this line with the theoretical
curve, we read out the CEP.  For these three shots we found their
CEP values are 129$^{\circ}$, 279$^{\circ}$ and 16$^{\circ}$,
respectively, with an error of about 3$^{\circ}$. No effort was made
to optimize the choices of the three energy points.

To illustrate that the CEP of each laser shot indeed varies
randomly, we show the retrieved CEP for shot numbers from 1000 to
1050 in Fig.~3. Clearly the CEP varies randomly from shot to shot.

\begin{figure}
\mbox{\rotatebox{270}{\myscaleboxa{
\includegraphics{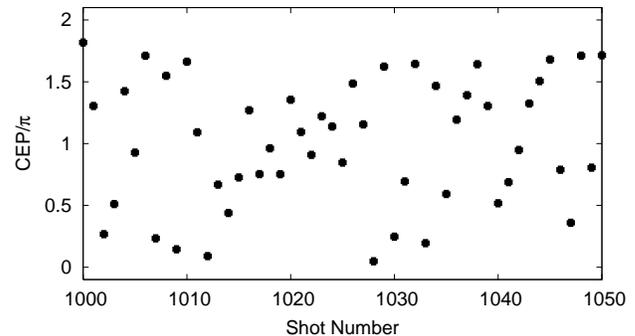}}}}
\caption{Absolute CEP extracted for experimental measurements of
Wittmann \emph{et al}.~\cite{Wittmann} from shot number 1000 to
1050.}
\end{figure}

The present method can also be used to determine the CEP of the
``phase-stabilized" laser pulses. We have electron momentum spectra
from Kling \emph{et al}.~\cite{Kling-private}. From their HATI
electron momentum spectra,  we deduced that the peak intensity is
$3.0\times10^{13}$ W/cm$^2$ and the pulse duration is 5.0~fs. The
mean wavelength of the laser used was 738 nm.  In Fig.~4(a), we show
the parametric plots  using ($E_1, E_2, E_3$)= (10.0, 13.0, 20.0) eV
for the 19 measured points. The ``size" of the ellipse is about the
same as in Fig.~2 since the pulse duration of 5.0 fs is close to the
4.7~fs in Fig.~2. However, the scattering of the experimental data
points from the theoretical ellipse is much larger for these
``phase-stabilized" laser pulses. This large scattering reflects the
lack of good phase stabilization. As shown in Wittmann \emph{et
al}., performing single-shot measurements on these
``phase-stabilized" pulses shows shot-to-shot CEP variations up to
about 20$^{\circ}$. Using the present method to retrieve the CEP, we
found that the retrieved CEP depends more sensitively on the values
of ($E_1,E_2,E_3$) used in the analysis, see Fig.~4(b), where two
sets of energy values were used to obtain the CEP for each
measurement. The straightline in Fig.~4(b) was drawn so that the
line best fits the deduced CEP's from successive measurements. On
the whole the phase decreases as the ``shot number" increases, thus
the phase has been stabilized at least partially.

\begin{figure}
\mbox{\rotatebox{270}{\myscaleboxa{
\includegraphics{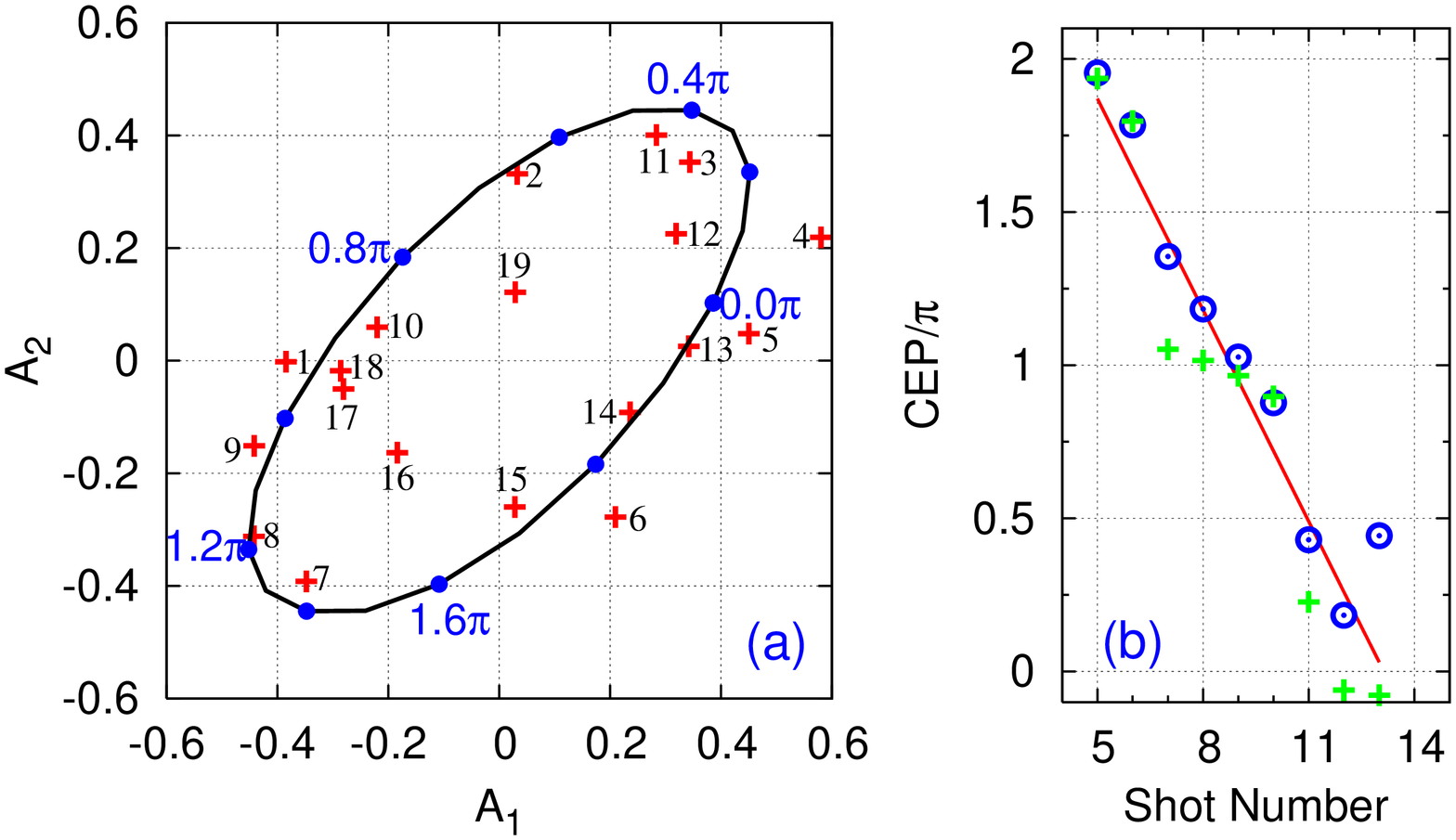}}}}
\mbox{\rotatebox{270}{\myscaleboxa{
\includegraphics{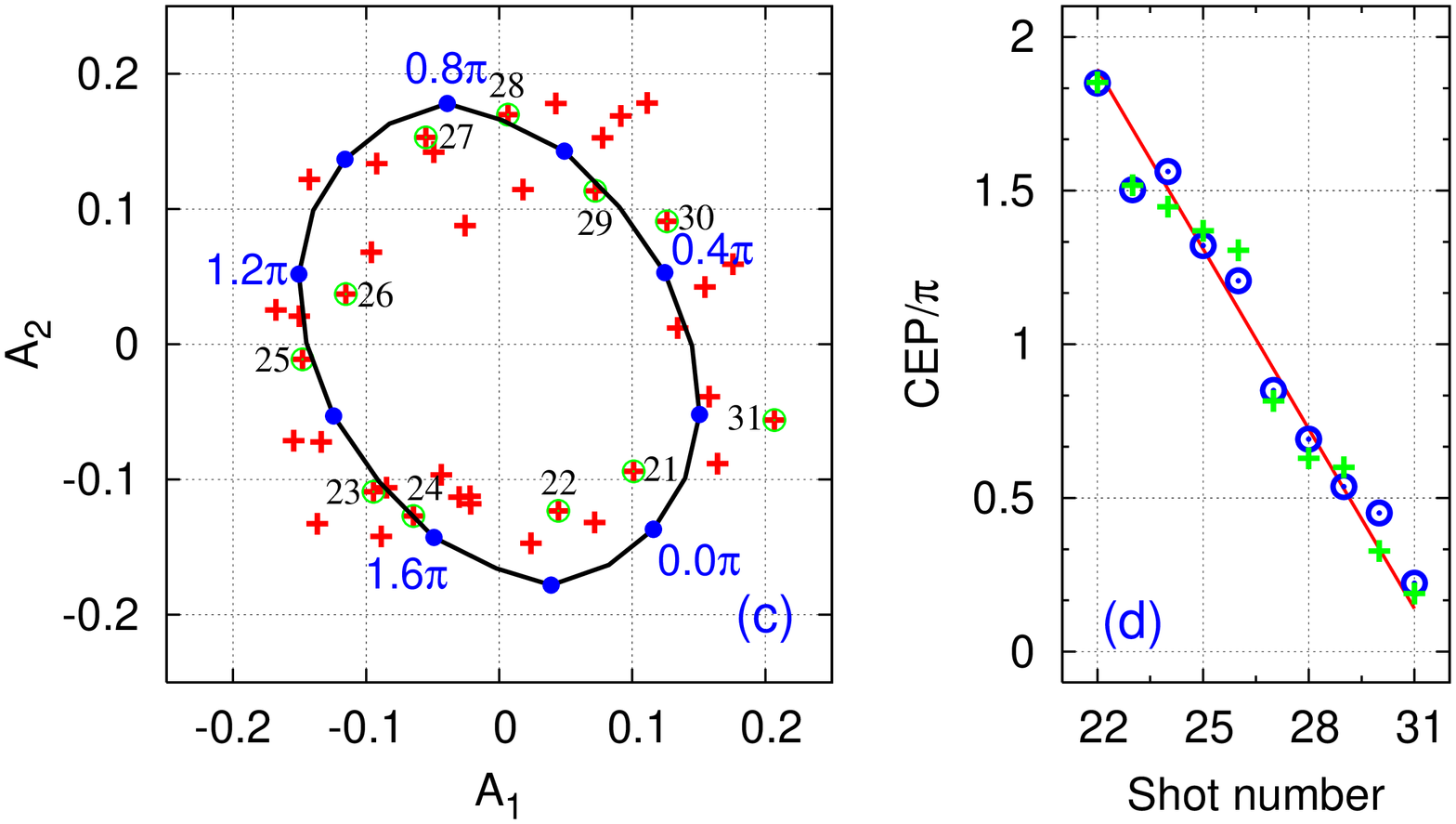}}}}
\caption{(Color online) Retrieval of CEP for ``sequential"
``phase-stabilized" measurements. (a) Comparison of asymmetries from
experiment (crosses) with asymmetry ellipse from theory (solid
line). The asymmetries are calculated using ($E_1,E_2,E_3$)=(10.0,
13.0, 20.0) eV. Experimental data for 19 shots are from Kling
\emph{et al}.~\cite{Kling-private} using wavelength of 738~nm.   The
peak intensity in the theoretical simulation is $3.0\times 10^{13}$
W/cm$^2$ and the pulse duration is 5.0~fs. (b) The absolute CEP's
for some shots extracted from (a) (circles), compared to those
extracted using ($E_1,E_2,E_3$)=(10.0, 15.0, 20.0) eV (crosses).
(c,d) Similar comparison for "phase-stabilized" long pulse
measurements from Kling \emph{et al}.~\cite{Kling08}. The data
consist of 40 sequential shots with wavelength of 760~nm. In (c),
shot numbers from 21 to 31 are marked. The peak intensity used in
the theory simulation is $1.05\times 10^{14}$ W/cm$^2$ and the pulse
duration is 7.0~fs. The energies used for the asymmetry calculations
are ($E_1,E_2,E_3$)=(40.0, 50.0, 60.0) eV and the retrieved phases
are shown as circles in (d). For the crosses in (d) the energies
used are (40.0, 45.0, 55.0) eV. The straightlines in (b) and (d) are
best fits to the deduced CEP's.}
\end{figure}

The determination of CEP is easier for shorter pulses. Previously,
Kling \emph{et al}.~\cite{Kling08} reported HATI electron momentum
spectra for longer pulses. Their data were analyzed by Micheau
\emph{et al}.~\cite{micheau} earlier, also using the QRS, but the
method assumed that CEP differences between successive measurements
are constant. For this same set of data, Micheau \emph{et al}. found
a pulse length of 6.7~fs and peak intensity of $1.05\times10^{14}$
W/cm$^2$. Using the present method we analyzed the 40 ``shots" from
Kling \emph{et al}.~\cite{Kling08}, and found that the pulse length
is 7.0 fs and the peak intensity is $1.05\times10^{14}$ W/cm$^2$,
close to the values from Micheau \emph{et al}.~\cite{micheau}. For
these longer pulses, the asymmetries $A_1$ and $A_2$ are much
smaller, see Fig.~4(c). Due to the remaining jittering of the phase
stabilized pulses from shot to shot, the scattering of the
experimental asymmetries around the theory ellipse are much larger.
For the ten measurements over the 2$\pi$ range, we notice again that
the retrieved CEP's depend more sensitively on the ($E_1,E_2,E_3$)
used for the retrieval, see Fig.~4(d), similarly to what was
observed in Fig.~4(b).

In summary, using the recently developed quantitative rescattering
theory, we propose an all-non-optical method to efficiently retrieve
accurate peak laser intensity, pulse duration and the CEP of each
single laser shot from the measured high-energy ATI electron
spectra. The method is very robust. The computational effort is very
small and requires no iterations. Unlike Wittmann \emph{et
al}.~\cite{Wittmann}, we show that there is no need to retrieve the
peak intensity and pulse duration using the optical method. In the
future, CEP dependence measurements can be carried out using
non-phase-stabilized lasers. By simple CEP tagging, the strong-field
effects on the waveform of ultrashort pulses can be accurately
carried out to achieve temporal resolution of a few attoseconds. The
present all-non-optical method  also characterizes the laser pulses
in the interaction region directly and avoid errors introduced by
the propagation of the laser beam.
\section{Acknowledgment}

We thank Matthias Kling and Marc Vrakking for providing high-energy
electron momentum spectra in \cite{Kling-private} for analysis, as
shown in Figs.~4(a,b). CDL also acknowledges earlier discussions on
the single-shot measurements with Gerhard Paulus. This work was
supported in part by Chemical Sciences, Geosciences and Biosciences
Division, Office of Basic Energy Sciences, Office of Science, US
Department of Energy.

\end{document}